%%%%%%%%%%%%%%%%%%%%%%%%%%%%%%%%%%%%%%%%%%%%%%%%%%%%%%%%%%%%%%%%%%%%%%%%%%%%%%
% Unified first law of black-hole dynamics and relativistic thermodynamics
% Sean A. Hayward
%%%%%%%%%%%%%%%%%%%%%%%%%%%%%%%%%%%%%%%%%%%%%%%%%%%%%%%%%%%%%%%%%%%%%%%%%%%%%%

\font\lbf=cmbx10 scaled\magstep1

\def\bs{\bigskip}
\def\ms{\medskip}
\def\np{\vfill\eject}

\def\ni{\noindent}
\def\cl{\centerline}

\def\title#1{\cl{\lbf #1}}
\def\ref#1#2#3#4{#1\ {\it#2\ }{\bf#3\ }#4\par}
\def\refb#1#2#3{#1\ {\it#2\ }#3\par}
\def\CQG{Class.\ Qu.\ Grav.}
\def\PR{Phys.\ Rev.}
\def\PRL{Phys.\ Rev.\ Lett.}
\def\PTP{Prog.\ Theor.\ Phys.}

\def\d{\hbox{d}}
\def\e{\hbox{e}}
\def\p{\partial}
\def\S{S_\circ}
\def\E{{\cal E}}
\def\f#1#2{{\textstyle{#1\over#2}}}

\magnification=\magstep1

\title{Unified first law of black-hole dynamics and relativistic thermodynamics}
\bs\cl{\bf Sean A. Hayward}
\ms\cl{Yukawa Institute for Theoretical Physics, Kyoto University, 
Kyoto 606-8502, Japan}
\ms\cl{\tt hayward@yukawa.kyoto-u.ac.jp}
\bs\ni{\bf Abstract.}
A unified first law of black-hole dynamics and relativistic thermodynamics 
is derived in spherically symmetric general relativity.
This equation expresses the gradient of the active gravitational energy $E$
according to the Einstein equation, divided into energy-supply and work terms.
Projecting the equation along the flow of thermodynamic matter 
and along the trapping horizon of a black hole yield, respectively, 
first laws of relativistic thermodynamics and black-hole dynamics.
In the black-hole case, 
this first law has the same form as the first law of black-hole statics,
with static perturbations replaced by the derivative along the horizon.
In particular, 
there is the expected term involving the area and surface gravity, 
where the dynamic surface gravity is defined by 
substituting the Kodama vector and trapping horizon 
for the Killing vector and Killing horizon 
in the standard definition of static surface gravity.
The remaining work term is consistent with, for instance, 
electromagnetic work in special relativity.
The dynamic surface gravity vanishes for degenerate trapping horizons 
and satisfies certain inequalities involving the area and energy 
which have the same form as for stationary black holes.
Turning to the thermodynamic case, the quasi-local first law has the same form,
apart from a relativistic factor, as the classical first law of thermodynamics,
involving heat supply and hydrodynamic work, 
but with $E$ replacing the internal energy.
Expanding $E$ in the Newtonian limit shows that 
it incorporates the Newtonian mass, kinetic energy, 
gravitational potential energy and thermal energy 
(internal energy with fixed zero).
There is also a weak type of unified zeroth law: 
a Gibbs-like definition of thermal equilibrium 
requires constancy of an effective temperature,
generalising the Tolman condition and the particular case of Hawking radiation,
while gravithermal equilibrium further requires constancy of surface gravity.
Finally, it is suggested that 
the energy operator of spherically symmetric quantum gravity is determined by 
the Kodama vector, which encodes a dynamic time related to $E$.
\bs\cl{PACS: 04.70.-s, 04.20.-q, 05.70.-a}
\bs\cl{Revised 5th March 1998}
\np\ni
{\bf I. Introduction}
\ms\ni
Analogies between the laws of thermodynamics 
and certain properties of black holes have aroused great interest,
particularly since Hawking showed that 
the analogy becomes actual in at least one respect, 
namely that the surface gravity determines the effective temperature 
of a quantum field on a stationary black-hole background.
This unexpected link between two hitherto unrelated branches of physics
has inspired the hope that they may be unified.
It is generally thought that there should be some deeper explanation 
involving a quantum theory of gravity.
However, this work reveals a purely classical connection.

Understanding this requires a paradigm shift 
from black-hole statics to black-hole dynamics,
meaning a theory of dynamically evolving black holes 
rather than just stationary black holes (and perturbations thereof). 
This is analogous to the shift from thermostatics to thermodynamics,
meaning the theory of dynamically evolving thermal systems 
rather than just those in thermal equilibrium (and perturbations thereof). 
Many researchers apparently still believe that 
the relevant generalisation concerns event horizons,
following the textbook definition of black holes by event horizons.
However, this definition makes essential use of the global assumption 
of asymptotic flatness.
The real universe is thought not to be asymptotically flat,
so that event horizons do not actually exist.
Moreover, even theoretically, the global nature of event horizons means that 
they cannot be physically located by observers.
Their location is unknown until the universe has ended.
For the same reason, event horizons have no local dynamical significance.
An event horizon might be passing through the reader right now,
without any physical effect.
In short, it seems that event horizons are not the relevant generalisation.
According to this view, the relevant second law of black-hole dynamics
cannot be what the textbooks state, 
namely Hawking's result that the area of an event horizon is non-decreasing.
This is reflected by the absence of a corresponding first law 
for event horizons.
The textbook first and zeroth laws concern stationary black holes,
so are more accurately described as laws of black-hole statics.
The textbook first and second laws do not even involve the same derivative,
thereby undermining the analogy with thermodynamics.

A new framework for black holes was introduced in a previous paper [1], 
referred to as Paper~I.
This dispensed with global assumptions and proposed a general, dynamical, quasi-local definition of black holes in terms of {\sl trapping horizons},
hypersurfaces foliated by marginal surfaces.
Paper~I also derived a second law of black-hole dynamics,
expressing the increase of area element along the trapping horizon.
This should not be confused with Hawking's second law for event horizons.
In the new framework, 
black-hole dynamics is formulated as the dynamics of trapping horizons.

A corresponding first law of black-hole dynamics requires 
a definition of surface gravity for non-stationary black holes.
This article gives a new definition of surface gravity 
for spherically symmetric black holes, 
with the same form as the usual stationary definition 
involving the Killing vector on the Killing horizon,
but instead using the Kodama vector on the trapping horizon.
The Kodama vector encodes a preferred time in spherically symmetric space-times 
and has physical properties analogous to those of Killing vectors.

Another key concept is energy.
In general, there is no agreed definition of gravitational energy 
in the theory of general relativity.
However, in spherical symmetry there is the Misner-Sharp energy $E$,
which does have all the physical properties one might expect of 
active gravitational energy, as explained in another recent paper [2],
referred to as Paper~II.
In particular, $E$ has the expected behaviour in various limits:
vacuum, small-sphere, large-sphere, Newtonian, test-particle 
and special-relativistic limits.
It is also intimately related to the characteristic strong-gravity phenomena,
namely black holes and singularities.

The first law of black-hole dynamics given here 
just expresses the derivative of $E$ along the trapping horizon, 
according to the Einstein equation.
This turns out to involve the area and surface gravity as expected,
with additional terms depending on the matter model,
such as an electromagnetic term which has the form of electromagnetic work.
This first law has the same form as the first law of black-hole statics,
with the perturbation being replaced by the derivative along the horizon.
In comparison to most references, 
it should be noted that this involves the energy on the horizon,
rather than the energy at infinity.
The new first and second laws involve the same derivative,
the dynamical derivative along the trapping horizon.

The next observation is that, for a perfect fluid, 
the derivative of $E$ along the fluid flow 
is given by the hydrodynamic work, as shown in Paper~II.
This therefore reads as a relativistic version of 
the classical first law of thermodynamics, 
albeit for the thermodynamically trivial case with no thermal flux.
To include such thermodynamic effects, 
it is necessary to generalise the model of a perfect fluid.
A generally relativistic theory of thermodynamics has been developed 
in an accompanying paper [3],
referred to as Paper~III.
This theory models heat relativistically by an energy tensor,
called the thermal energy tensor,
whose various components are thermal energy density, flux and stress.
(Thermal energy is internal energy with fixed zero).
This turns out to add a heat-supply term to the first law,
consistent with the classical first law of thermodynamics. 
Actually, this requires no thermodynamics 
other than the form of the energy tensor given by Eckart [4]
and the definition of heat supply as an integral of thermal flux.
Thus we have a quasi-local first law of relativistic thermodynamics.
This shows that the gravitational energy $E$ incorporates the thermal energy,
as may be verified explicitly in the Newtonian limit.
Then the unified first law is just the equation expressing the gradient of $E$
according to the Einstein equation. 
Projecting the unified first law along the material flow 
or along a trapping horizon yields the first law of thermodynamics 
or black-hole dynamics, respectively.

One might also expect there to be a unified zeroth law, 
concerning gravithermal equilibrium.
The Gibbs definition of thermal equilibrium is constancy of 
the partial derivative of entropy with respect to thermal energy.
Replacing thermal energy with gravitational energy $E$ in this definition 
yields constancy of an effective temperature,
differing from the actual temperature by a red-shift factor.
This generalises the Tolman condition [6] for gravithermal equilibrium
and agrees with the case of Hawking radiation [5] on a stationary background.
Surface gravity is also constant for stationary black holes,
thereby linking the two zeroth laws.

The article is organised as follows.
Section~II reviews the relevant properties of spherically symmetric space-times,
defining $E$ and deriving the unified first law.
Section~III considers work in the case of electromagnetism 
and Section~IV reviews black-hole statics.
Section~V describes the Kodama vector or dynamic time,
Section~VI introduces the dynamic surface gravity 
and Section~VII derives the first law of black-hole dynamics.
Section~VIII derives the corresponding first law of relativistic thermodynamics
and shows how $E$ incorporates thermal energy.
Section~IX considers thermal equilibrium, gravithermal equilibrium 
and the zeroth laws.
Section~X concludes.
\bs\ni
{\bf II. Unified first law}
\ms\ni
In spherical symmetry, 
the area $A$ of the spheres of symmetry is a geometrical invariant.
It is convenient to use the areal radius $r=\sqrt{A/4\pi}$, giving 
$$A=4\pi r^2.\eqno(2.1)$$
Following Papers I and II, 
a sphere is said to be {\sl untrapped, marginal or trapped}
as $\nabla^\sharp r$ is spatial, null or temporal respectively,
$\nabla^\sharp r=0$ being degenerate.
Here $\nabla$ denotes the covariant derivative operator
and $\sharp$ denotes the contravariant dual (index raising)
with respect to the space-time metric $g$.
Similarly, $\flat$ will denote the covariant dual (index lowering).
If the space-time is time-orientable 
and $\nabla^\sharp r$ is future (respectively past) causal,
then the sphere is said to be future (respectively past) trapped or marginal.
Untrapped or marginal spheres have a spatial orientation 
given by the direction of $\nabla^\sharp r$:
a spatial or null direction is said to be 
{\sl outward} (respectively {\sl inward}) 
if $r$ is increasing (respectively decreasing) in that direction.
A hypersurface foliated by marginal spheres is called a {\sl trapping horizon}.
A trapping horizon is said to be {\sl outer, degenerate} or {\sl inner} 
as $\nabla^2r>0$, $\nabla^2r=0$ or $\nabla^2r<0$ respectively.

Paper~II proposed that future (respectively past) outer trapping horizons 
be taken as the dynamical definition of the outer boundaries of 
black (respectively white) holes. 
Black holes are thereby characterised quasi-locally,
actually locally in spherical symmetry.
This definition should not be confused with 
those of event horizons or apparent horizons [7],
both of which require the global assumption of asymptotic flatness. 
It is also unnecessary to make assumptions about 
the global nature of the trapping horizon, 
for instance that it remains of the future outer type.
Such questions are still of interest, 
but depend on the dynamics of the chosen matter model.
The results of this paper are essentially local, 
independent of the matter model except for energy conditions and,
with minor exceptions for degenerate cases, apply to any trapping horizon.

The Misner-Sharp energy [8] may be defined by
$$E=\f12r\left(1-\nabla r\cdot\nabla^\sharp r\right)\eqno(2.2)$$
where the dot denotes contraction,
the sign convention is that spatial metrics are positive definite
and the Newtonian gravitational constant is unity.
Paper~II investigated the physical properties of $E$ in detail,
establishing that it represents active gravitational energy.
This point deserves emphasis, 
since the literature contains many definitions of so-called energy
which do not have the relevant properties in physically understood limits;
a list of references may be found in [9].
In comparison, $E$ does behave as active gravitational energy
in the vacuum, small-sphere, large-sphere, Newtonian, test-particle 
and special-relativistic limits.

Two invariants of the (contravariant) energy tensor $T$ will be useful:
the function
$$w=-\f12\hbox{trace}\,T\eqno(2.3)$$
and the vector
$$\psi=T\cdot\nabla r+w\nabla^\sharp r\eqno(2.4)$$
where the trace refers to the two-dimensional space 
normal to the spheres of symmetry.
It is also convenient to use the areal volume 
$$V=\f43\pi r^3\eqno(2.5)$$ 
due to the relation
$$\nabla V=A\nabla r.\eqno(2.6)$$ 
Then the gradient of $E$ is determined by the Einstein equations as
$$\nabla E=A\psi^\flat+w\nabla V.\eqno(2.7)$$
This equation turns out to be the {\sl unified first law}, 
as explained subsequently.
This may be generalised beyond spherical symmetry 
in terms of the Hawking energy [10].

One may physically interpret $w$ as an energy density 
and $\psi$ as an energy flux or momentum density.
Actually, assuming the null energy condition, 
$w$ is a lower bound for the energy density measured by an observer.
Assuming the dominant energy condition, $w\ge0$.
Assuming the null energy condition, 
$\psi$ is past (respectively future) causal 
in future (respectively past) trapped regions,
and outward spatial or null in untrapped regions.
If the space-time is asymptotically flat, 
one finds that $\psi$ becomes future (respectively past) null 
at future (respectively past) null infinity, reducing to the Bondi flux.
This can be seen from the first law (2.7), 
since $w\nabla V$ tends to zero at infinity.
Thus $\psi$ is a local version of the Bondi flux,
actually the outward flux minus the inward flux.
Then the first term $A\psi$ in the unified first law may be interpreted as 
an energy-supply term, analogous to the heat-supply term 
in the classical first law of thermodynamics,
while the second term $w\nabla V$ may be interpreted as a work term.

To check the above properties,
one may use double-null coordinates $(\xi^+,\xi^-)$,
in terms of which the line-element is
$$\d s^2=r^2\d\Omega^2+2g_{+-}\d\xi^+\d\xi^-\eqno(2.8)$$
where $\d\Omega^2$ refers to the unit sphere
and $r\ge0$ and $g_{+-}<0$ are functions of $\xi^\pm$.
Then the coordinate forms of (2.2--4) and (2.7) are
$$\eqalignno
{&E=r\left(\f12-g^{+-}\p_+r\p_-r\right)&(2.9a)\cr
&w=-g_{+-}T^{+-}&(2.9b)\cr
&\psi=T^{++}\p_+r\p_++T^{--}\p_-r\p_-&(2.9c)\cr
&\p_\pm E=A\psi_\pm+w\p_\pm V
=-Ag^{+-}\left(T_{+-}\p_\pm r-T_{\pm\pm}\p_\mp r\right).&(2.9d)\cr}$$
The last expression was derived from the Einstein equations in Paper~II.
In terms of these coordinates, a sphere is trapped if $\p_+r\p_-r>0$,
being future trapped if $\p_\pm r<0$ and past trapped if $\p_\pm r>0$,
taking $\p_\pm$ to be future pointing.
In an untrapped region, $\p_+r\p_-r<0$ 
and one may locally choose the orientation $\p_+r>0$, $\p_-r<0$, 
meaning that $\p_+$ is outward and $\p_-$ is inward.
As in Papers I and II, 
note that the null energy condition requires $T_{\pm\pm}\ge0$
and that the dominant energy condition requires $T_{+-}\ge0$.
The Einstein equations translated from Paper~II are
$$\eqalignno
{&\p_\pm\p_\pm r-\p_\pm\log(-g_{+-})\p_\pm r=-4\pi rT_{\pm\pm}&(2.10a)\cr
&r\p_+\p_-r+\p_+r\p_-r-\f12 g_{+-}=4\pi r^2T_{+-}&(2.10b)\cr
&r^2\p_+\p_-\log(-g_{+-})-2\p_+r\p_-r+g_{+-}
=8\pi r^2(g_{+-}T_\theta^\theta-T_{+-})&(2.10c)\cr}$$
where $\theta$ is a standard latitude.
\bs\ni
{\bf III. Electromagnetic work}
\ms\ni
Since the desired first law of black-hole dynamics will involve work terms,
it is necessary to identify the correct definition of work 
for the relevant matter fields.
This is done as follows for the Maxwell electromagnetic field 
(excluding magnetic monopoles).
The only non-zero components of a spherically symmetric
electromagnetic field tensor $F$ are given by
$$\E=-g^{+-}F_{+-}=g^{+-}F_{-+}\eqno(3.1)$$
which may be intepreted as the electric field strength,
the magnetic field vanishing.
A standard definition of charge [11] is
$$e={1\over{4\pi}}\oint\E=r^2\E\eqno(3.2)$$
where the integral is over a sphere of symmetry, the area form being implicit.
Thus the field may be characterised by either $\E$ or $e$.
The electromagnetic energy tensor $T_e$ has one independent component,
yielding
$$\eqalignno
{&w_e=\E^2/8\pi&(3.3a)\cr&\psi_e=0&(3.3b)\cr}$$
where the notation is that of (2.3--4) 
with the subscript $e$ referring to the electromagnetic field.
Therefore the unified first law (2.7) reads
$$\nabla E={\E^2\nabla V\over{8\pi}}+w_o\nabla V+A\psi_o^\flat\eqno(3.4)$$
where the subscript $o$ refers to other (non-electromagnetic) fields,
i.e.\ $T=T_e+T_o$ and so on.
In the absence of other fields, the Reissner-Nordstr\"om case is recovered.
This identifies the electromagnetic work as
$\E^2\nabla V/8\pi$.
This agrees with the standard expression for electric work [12]
in special relativity, which is
$$W={1\over{8\pi}}\int_\Sigma\E^2\d V\eqno(3.5)$$
where $\d V$ is the volume form of a flat spatial hypersurface 
containing the region $\Sigma$.
The integrand is just the local energy density $w_e$ 
of the electromagnetic field.
Some authors accept this expression for work only for infinite $\Sigma$,
allowing it to be rewritten in various ways 
involving boundary terms which vanish at infinity.
However, such terms are generally non-zero at a regular centre.
Indeed, typical suggestions become infinite.

It is also possible to write the electromagnetic work 
in terms of the electric field covector 
$$\hat\E=F\cdot k\eqno(3.6)$$
where $k$ is the Kodama vector defined subsequently by (5.1).
Then $\hat\E=\E\nabla r$, so that the electromagnetic work is
$${\E^2\nabla V\over{8\pi}}=\f12e\hat\E.\eqno(3.7)$$
The second expression has the same form as the work $e_0\hat\E$ 
done on a test charge $e_0$ 
by the electric field $\hat\E$ in special relativity, 
with the half arising to avoid counting the self-interaction twice.
In other words, the work is that done on the charge distribution 
by its own electric field.
\bs\ni
{\bf IV. First law of black-hole statics}
\ms\ni
This section reviews the theory of black-hole statics, 
or equilibrium mechanics, 
which concerns stationary black holes;
see e.g.\ Carter [13] or Wald [14].
Stationary space-times are defined by a Killing vector $\xi$ 
generating asymptotic time translations.
That is, $\xi$ satisfies the Killing equation
$$\nabla\otimes\xi^\flat=0\eqno(4.1)$$
where $\otimes$ denotes the symmetric tensor product.
A Killing horizon is a null hypersurface such that 
$\xi$ is tangent to the null generators.
Killing horizons define the boundaries of stationary black holes, 
white holes and cosmological regions.
It follows from the definition that 
$\xi\cdot(\nabla\wedge\xi^\flat)$ is parallel to $\xi^\flat$ 
on a Killing horizon, 
where $\wedge$ denotes the antisymmetric tensor product.
The proportionality constant defines the surface gravity $\kappa$:
$$\xi\cdot(\nabla\wedge\xi^\flat)=\kappa\xi^\flat
\qquad\hbox{on a Killing horizon.}\eqno(4.2)$$
Taking the standard example of the source-free electromagnetic field,
the only spherically symmetric solution to the Einstein-Maxwell equations
is the Reissner-Nordstr\"om solution,
parametrised by two constants, the charge $e$ and asymptotic energy $m$;
these provide an analogue of thermostatic state space 
for stationary black holes.
For $m\ge|e|$ the solution describes a black hole 
with surface gravity $\kappa=r^{-2}\sqrt{m^2-e^2}$
on the Killing horizon at radius $r=m+\sqrt{m^2-e^2}$.
Evaluating these quantities on the horizon yields
$$\d m={\kappa\d A\over{8\pi}}+{e\d e\over{r}}\eqno(4.3)$$
where $\d$ denotes the differential in state space $(m,e)$.
This is the form usually given for the first law in this case,
though it may be generalised to non-stationary perturbations [15]
(of stationary solutions).
However, this form is not suitable for generalisation 
to space-times which are not asymptotically flat, 
due to the appearance of the asymptotic energy $m$.
Instead, one may use the quasi-local energy $E$,
which for the Reissner-Nordstr\"om solution is 
$$E=m-{e^2\over{2r}}.\eqno(4.4)$$
Rewriting (4.3) in terms of $E$ yields
$$\d E={\kappa\d A+\E^2\d V\over{8\pi}}\eqno(4.5)$$
where $\E=e/r^2$ as in (3.2).
This is a physically acceptable form for the first law 
of black-hole statics in the electromagnetic case,
since the last term may be interpreted as electromagnetic work,
in agreement with special relativity (3.5)
and the general expression (3.7) for electromagnetic work.
Thus the analogue of internal energy for a black hole
is the energy $E$ on the horizon, rather than the energy at infinity;
put another way, 
the energy of the black hole rather than the energy of the space-time.
This is exactly as one would have expected by physical intuition
and provides further evidence for the physical correctness of $E$ 
as a definition of active gravitational energy.
\bs\ni
{\bf V. Dynamic time}
\ms\ni
In black-hole dynamics, unlike black-hole statics, 
there is no Killing vector to define such quantities as surface gravity.
However, in spherical symmetry there is a natural analogue, the vector
$$k=\hbox{curl}\,r\eqno(5.1)$$
where the curl refers to the two-dimensional space 
normal to the spheres of symmetry.
In double-null coordinates,
$$k=-g^{+-}\left(\p_+r\p_--\p_-r\p_+\right)\eqno(5.2)$$
up to orientation, which in an untrapped region may be locally fixed 
such that $k$ is future-pointing.
Kodama [16] introduced $k$ and Paper~II discussed some properties.
It follows that
$$\eqalignno{&k\cdot\nabla r=0&(5.3a)\cr
&k\cdot k^\flat={2E\over{r}}-1&(5.3b)\cr}$$
which equivalently defines $k$.
So $k$ is spatial, null or temporal 
for trapped, marginal or untrapped spheres respectively.
In particular, the definition of the boundary of stationary black holes, 
a hypersurface where the Killing vector is null, 
may be generalised to a hypersurface where the Kodama vector is null,
which is equivalent to the definition of trapping horizon. 
Unlike a Killing horizon, a trapping horizon is generally not null.

The Kodama and Killing vectors agree for the Reissner-Nordstr\"om black hole.
More generally, they agree in a stationary, spherically symmetric space-time
if $k$ commutes with $\nabla^\sharp r$.
This condition implies that 
one may introduce coordinates $(r,t)$ where $k=\p/\p t$,
for which the line-element becomes
$$\d s^2=r^2\d\Omega^2+\left(1-{2E\over{r}}\right)^{-1}\d r^2
-\left(1-{2E\over{r}}\right)\d t^2.\eqno(5.4)$$
This is a Schwarzschild-like form, with $E$ generally a function of $(r,t)$.
Then examine the various components of the Killing equation (4.1):
$$\xi^c\p_cg_{ab}+2g_{c(a}\p_{b)}\xi^c=0\eqno(5.5)$$
for a Killing vector $\xi$ normal to the spheres of symmetry.
The components tangential to the spheres of symmetry yield $\xi\cdot\nabla r=0$,
so that $\xi$ is parallel to $k$: $\xi=\xi^0k$.
The $rr$ component yields $\xi\cdot\nabla E=0$,
the $tt$ component yields $k\cdot\nabla\xi^0=0$,
and the $tr$ component yields $\nabla^\sharp r\cdot\nabla\xi^0=0$,
so that $\xi^0$ is constant.
If the space-time is asymptotically flat 
and the Killing vector is normalised as usual by 
$\xi\cdot\xi^\flat\to-1$ as $r\to\infty$,
then $\xi^0=1$ since $k\cdot k^\flat\to-1$ as $r\to\infty$.
Thus the Kodama vector reduces to the Killing vector, $\xi=k$.

Another physically relevant property of $k$ is that 
it can be used to define $E$ as a Noether charge, as shown in Paper~II.
Quoting this result, both $k$ and the corresponding energy-momentum density 
$$j=-T\cdot k^\flat\eqno(5.6)$$
are conserved: 
$$\eqalignno
{&\nabla\cdot k=0&(5.7a)\cr &\nabla\cdot j=0.&(5.7b)\cr}$$
Therefore the Gauss theorem yields conserved charges 
which turn out to be just the areal volume and energy:
$$\eqalignno
{&V=-\int_\Sigma{*}u^\flat\cdot k&(5.8a)\cr
&E=-\int_\Sigma{*}u^\flat\cdot j
=\int_\Sigma{*}u\cdot T^\flat\cdot k&(5.8b)\cr}$$
where $*$ denotes the volume form and $u$ the unit future normal vector
of an arbitrary spatial hypersurface $\Sigma$ with regular centre.
Paper~II also showed how the H\'aj\'\i\v cek [17] energy $\tilde E$ 
of a spherical shell of test particles of rest mass $m$ 
and velocity (unit future-temporal vector) $u$ may be defined analogously by
$$\tilde E=\int_\Sigma{*}u\cdot\tilde T^\flat\cdot k\eqno(5.9)$$
where the energy tensor $\tilde T$ of the shell is
$$\tilde T=m\delta u\otimes u\eqno(5.10)$$
and $\delta$ is the Dirac distribution with support on the shell.
Then 
$$\tilde E=-mu^\flat\cdot k.\eqno(5.11)$$
This has the same form as the standard definition of energy 
$-mu^\flat\cdot\xi$ for test particles in the stationary case [14].

This allows a definition of {\sl ergoregion} as a region where 
test particles may have positive or negative energy $\tilde E$ at each point, 
depending on their velocity $u$ [18].
The overall sign of $\tilde E$ is conventional, as in the stationary case, 
with $\tilde E>0$ for future pointing $k$ in untrapped regions.
Ergoregions are physically important because they allow energy extraction 
by the Penrose process [14], 
whereby discarding negative-energy particles yields a net gain in energy.
It follows from (5.11) that ergoregions are regions where $k$ is spatial.
Again this is analogous to the stationary case.
In spherical symmetry, ergoregions therefore coincide with trapped regions.

In summary, the Kodama vector generates a preferred flow of time 
in spherically symmetric space-times,
analogous to the Killing vector of stationary space-times
and sharing its relevant physical properties.
Moreover, it generates active gravitational energy as a conserved charge,
suggesting the name dynamic time.
This further suggests that dynamic gravitational entropy might be defined 
by analogy with the method of Wald [15].
\bs\ni
{\bf VI. Dynamic surface gravity}
\ms\ni
The above properties suggest defining dynamic surface gravity 
by replacing the Killing horizon with the trapping horizon
and the Killing vector with the Kodama vector.
To see whether this is possible, 
note firstly that there appears to be an ambiguity in such an analogy,
since the Killing equation (4.1) allows the definition of surface gravity (4.2) 
to be rewritten using different linear combinations of 
$\xi\cdot(\nabla\wedge\xi^\flat)$ and $\xi\cdot(\nabla\otimes\xi^\flat)$.
So one needs to calculate both
$$\eqalignno
{&k\cdot(\nabla\wedge k^\flat)=\left(E/r^2-4\pi rw\right)\nabla r&(6.1a)\cr
&k\cdot(\nabla\otimes k^\flat)=4\pi r\psi^\flat.&(6.1b)\cr}$$
The term proportional to $\nabla r$ is what is needed, 
since $\nabla r=\pm k^\flat$ on a trapping horizon $\p_\pm r=0$, 
but the term proportional to $\psi$ is generally not parallel to $k$.
This resolves the ambiguity, 
leading to the definition of dynamic surface gravity
$$\kappa={E\over{r^2}}-4\pi rw.\eqno(6.2)$$
When $w=0$, this has the same form as Newtonian surface gravity,
or indeed Newtonian gravitational acceleration anywhere,
with $E$ replacing the Newtonian mass.
Explicitly, the definition satisfies
$$k\cdot(\nabla\wedge k^\flat)=\kappa\nabla r\eqno(6.3)$$
and therefore
$$k\cdot(\nabla\wedge k^\flat)=\pm\kappa k^\flat
\qquad\hbox{on a trapping horizon $\p_\pm r=0$.}\eqno(6.4)$$
This has the same form as 
the standard stationary definition of surface gravity (4.2).
The new definition therefore recovers the Reissner-Nordstr\"om surface gravity, 
but generally differs from the definitions of both Paper~I 
(evaluated in spherical symmetry) and Fodor et al.\ [19].
Based on the stationary case,
one might conjecture that a dynamic spherically symmetric quantum field 
has a local Hawking temperature $\hbar\kappa/2\pi$, 
but that remains an open question.

The dynamic surface gravity may equivalently be expressed as
$$\kappa=\f12\hbox{div}\,\hbox{grad}\,r\eqno(6.5)$$
where the divergence and gradient refer to the two-dimensional space 
normal to the spheres of symmetry.
This may be regarded as the more fundamental, purely geometrical definition,
with the form (6.2) following from the Einstein equations.
Then outer, degenerate and inner trapping horizons,
as defined in Section~II or in Papers I or II, 
respectively have $\kappa>0$, $\kappa=0$ and $\kappa<0$.
The overall sign is conventional, but this confirms the desired property that 
surface gravity should vanish for degenerate black holes.

The dynamic surface gravity satisfies certain inequalities
involving the area and energy, assuming the dominant energy condition.
Directly from the definition (6.2), 
$$A\kappa\le4\pi E.\eqno(6.6)$$
This holds anywhere in the space-time, 
though the interpretation of $\kappa$ as surface gravity 
is usually intended only on a horizon.
Denoting values on the trapping horizon by a subscripted zero
and recalling $E_0=\f12r_0$, 
$$\kappa_0\le\sqrt{{\pi\over{A_0}}}\eqno(6.7)$$
or equivalently $\kappa_0\le1/2r_0$.
This has the same form as an inequality of Visser [20] 
for stationary black holes.
Thus for a black hole of given area, 
the surface gravity has an upper bound.

Moreover, Paper~II established the Penrose inequality 
$$\sqrt{\pi A_0}\le4\pi E\eqno(6.8)$$
where $E$ is evaluated anywhere in the untrapped region 
outward from the trapping horizon. 
(Recall that 
the outward orientation is defined for spatial and null directions).
Therefore
$$A_0\kappa_0\le4\pi E\eqno(6.9)$$
under the same conditions.
If the untrapped region is asymptotically flat,
the same inequality holds for the asymptotic (Bondi and ADM) energies,
which are limits of $E$ at null and spatial infinity respectively.
This has the same form as an inequality of Heusler [21] 
for stationary black holes.
Thus the total surface gravity of a black hole 
provides a lower bound for the energy measured outside the black hole.
\bs\ni
{\bf VII. First law of black-hole dynamics}
\ms\ni
The desired first law of black-hole dynamics may now be given as
$$E'={\kappa A'\over{8\pi}}+wV'\eqno(7.1)$$
where the prime denotes the derivative along the trapping horizon,
$f'=z\cdot\nabla f$,
where $z$ is a vector tangent to the trapping horizon,
the normalisation of $z$ being irrelevant.
This follows by projecting the unified first law (2.7) along $z$, as follows.
The non-trivial part is to show that 
$A\psi^\flat\cdot z=\kappa A'/8\pi$.
Writing $z$ in double-null coordinates, $z=z^+\p_++z^-\p_-$,
and taking the horizon to be given by $\p_+r=0$, one has
$0=(\p_+r)'=z^+\p_+\p_+r+z^-\p_-\p_+r$,
which expands to
$0=(\p_+r)'=g_{+-}\kappa z^--4\pi rT_{++}z^+$,
using the Einstein equations (2.10) and the definition of $\kappa$ (6.2).
Therefore $A\psi^\flat\cdot z=Ag^{+-}T_{++}z^+\p_-r=r\kappa z^-\p_-r
=r\kappa r'=\kappa A'/8\pi$.

The term $\kappa A'/8\pi$ is exactly as expected 
for a first law of black-hole dynamics,
with the derivative along the horizon replacing the state-space differential 
of the first law of black-hole statics (4.5).
To complete the desired physical interpretation, 
the last term $wV'$ should be the work done along the horizon.
This can be seen for the electromagnetic field, for which the first law becomes
$$E'={\kappa A'+\E^2V'\over{8\pi}}+w_oV'\eqno(7.2)$$
as follows from (3.3).
The term $\E^2V'/8\pi$ 
is the electromagnetic work done along the horizon,
in agreement with the stationary (4.5) and special-relativistic (3.5) cases.
\bs\ni
{\bf VIII. Quasi-local first law of relativistic thermodynamics}
\ms\ni
Relativistic thermodynamics, as described for instance in Paper~III, 
involves a conserved material current $\rho u$, 
where $\rho$ is the density and $u$ the unit velocity vector, 
$$u\cdot u^\flat=-1\eqno(8.1)$$
with the physical interpretation as the flow vector of a fluid,
or as giving the centre-of-momentum frame of a solid.
The various components of the energy tensor $T$ of the matter define
the thermal energy density
$$\varepsilon=u\cdot T^\flat\cdot u-\rho\eqno(8.2)$$
the thermal flux
$$q=-\bot(T\cdot u^\flat)\eqno(8.3)$$
and the thermal stress 
$$\tau=\bot T\eqno(8.4)$$
where $\bot$ denotes projection by the spatial metric
$g+u^\flat\otimes u^\flat$, 
which is orthogonal to the material flow.
Therefore the energy tensor of the thermodynamic matter is
$$T=(\rho+\varepsilon)u\otimes u+2u\otimes q+\tau.\eqno(8.5)$$
Since specific thermal energy is $\varepsilon/\rho$,
this is effectively the energy tensor of Eckart [4],
who defined specific internal energy as $u\cdot T^\flat\cdot u/\rho$ 
plus an undetermined constant.

The vorticity $\bot(\nabla\wedge u^\flat)$ vanishes in spherical symmetry,
so there exist hypersurfaces orthogonal to the flow vector $u$,
labelled by a time coordinate $t$.
Take another coordinate $x$ on these hypersurfaces,
orthogonal to the spheres of symmetry.
Choosing vanishing shift vector, the line-element (2.8) becomes
$$\d s^2=r^2\d\Omega^2+\e^{2\zeta}\d x^2-\e^{2\phi}\d t^2\eqno(8.6)$$
where $(r,\phi,\zeta)$ are functions of $(t,x)$.
These are comoving coordinates adapted to the fluid.
The volume form of the hypersurfaces is given by 
$$\int_\Sigma{*}f=\int{f\p_xV\d x\over{\alpha}}\eqno(8.7)$$
where $\Sigma$ is a region of one of the hypersurfaces and
$$\alpha=\e^{-\zeta}\p_xr=\pm\left(1+\dot r^2-{2E\over{r}}\right)^{1/2}
\eqno(8.8)$$
where the dot denotes the material (or comoving) derivative, 
$\dot f=u\cdot\nabla f$, 
and the sign is that of $\p_xr$.
The second expression gives the definition of $E$ in comoving coordinates [8].
The red-shift factor $\alpha$ tends to one in the Newtonian limit. 

The unified first law (2.7), written in comoving coordinates, is
$$\eqalignno
{&\p_tE=A\e^{-2\zeta}(T_{tx}\p_xr-T_{xx}\p_tr)&(8.9a)\cr
&\p_xE=A\e^{-2\phi}(T_{tt}\p_xr-T_{tx}\p_tr).&(8.9b)\cr}$$
For the thermodynamic matter, the relevant energy components are
$$\eqalignno
{&T_{tt}=\e^{2\phi}(\rho+\varepsilon)&(8.10a)\cr
&T_{tx}=-\e^{\phi+\zeta}\hat q&(8.10b)\cr
&T_{xx}=\e^{2\zeta}p&(8.10c)\cr}$$
where $p=\tau^x_x$ is the radial pressure
and $\hat q$ is the strength of the thermal flux
$q^\flat=\hat q\nabla x/|\nabla x|$.
The heat supply as defined in Paper~III is
$$Q=-\int_{\hat\Sigma}\hat{*}\hat{q}\eqno(8.11)$$ 
where $\hat\Sigma$ is a region of a hypersurface generated by flowlines of $u$,
with volume form given by
$$\int_{\hat\Sigma}\hat{*}f=\int fA\e^\phi\d t.\eqno(8.12)$$ 
Thus
$$\Delta Q=-\e^\phi A\hat{q}\eqno(8.13)$$ 
where $\Delta f=\p_tf$.
Equivalently in spherical symmetry,
$\dot Q=-A\hat{q}$. 
Then the component (8.9a) of the unified first law along the flow reads
$$\Delta E=\alpha\Delta Q-p\Delta V.\eqno(8.14)$$
This is the desired quasi-local first law of relativistic thermodynamics.
Since the quantities involved are local in spherical symmetry, 
this first law may also be written in terms of the material derivative as 
$$\dot E=\alpha\dot Q-p\dot V.\eqno(8.15)$$
The first term is a relativistic modification involving $\alpha$
of the heat-supply term $\dot Q$
in the classical first law of thermodynamics.
The last term has the same form as the hydrodynamic work term 
of the classical case, 
with the relevant radial pressure $p$ 
and with the areal volume $V=\int_\Sigma{*}\alpha$ 
replacing the proper volume $\int_\Sigma{*}1$,
again differing by the relativistic factor $\alpha$.
Recalling the role of $E$ in black-hole dynamics, this establishes that 
the analogy between the two first laws is more than an analogy:
the two first laws are different projections of the unified first law.
Both express derivatives of the active gravitational energy 
according to the Einstein equation.

Unlike the classical first law of thermodynamics, 
which is an assumed principle,
the quasi-local first law has been derived from the Einstein equations 
and the material model.
Thus heat is included in general relativity as a form of energy,
the quasi-local first law being a consequence of 
the relativistic definition of energy.
Comparing with the non-relativistic first law, this indicates that 
the active gravitational energy $E$ must incorporate the thermal energy,
thereby linking thermodynamics and gravity.
This may be seen explicitly as follows.
The other component (8.9b) of the unified first law reads
$$\p_xE=(\rho+\varepsilon+\dot r\hat{q}/\alpha)\p_xV\eqno(8.16)$$
which integrates to
$$E=\int_\Sigma{*}(\alpha(\rho+\varepsilon)+\dot r\hat q).\eqno(8.17)$$
The integrand provides the definition of active gravitational energy density.
In this form, $E$ manifestly involves both thermal energy and thermal flux,
though the latter disappears in the Newtonian limit, as follows.

The Newtonian limit may be described as in Paper~II or III 
by inserting factors of the speed $c$ of light by the formal replacements
$$({*},r)\mapsto ({*},r)\qquad
\dot r\mapsto c^{-1}\dot r\qquad
\rho\mapsto c^{-2}\rho\qquad
(\varepsilon,\tau,E)\mapsto c^{-4}(\varepsilon,\tau,E)\qquad
q\mapsto c^{-5}q\eqno(8.18)$$
then taking the limit $c^{-1}\to0$.
Then
$$E=Mc^2+H+K+U+O(c^{-2})\eqno(8.19)$$
where the mass $M$, thermal energy or heat $H$, 
kinetic energy $K$ and gravitational potential energy $U$ are defined by
$$\eqalignno
{&M=\int_\Sigma{*}\rho&(8.20a)\cr
&H=\int_\Sigma{*}\varepsilon&(8.20b)\cr
&K=\int_\Sigma{*}\f12\rho\dot r^2&(8.20c)\cr
&U=-\int_\Sigma{*}{M\rho\over{r}}.&(8.20d)\cr}$$
All four quantities have the same form as in Newtonian theory.
Thus $E$ incorporates all the energies and mass present in Newtonian theory.
Again this confirms the correctness of $E$
as a definition of active gravitational energy.
\bs\ni
{\bf IX. Zeroth laws}
\ms\ni
The zeroth law of black-hole statics states that for a stationary black hole,
the surface gravity $\kappa$ is constant [13,14],
a trivial property in spherical symmetry.
Since a stationary space-time may be taken as the definition 
of gravitational equilibrium, this is analogous to 
the classical zeroth law of thermodynamics,
which states that the temperature is constant in thermal equilibrium.
In the Gibbs theory of thermostatics,
often called equilibrium thermodynamics or reversible thermodynamics, 
this is expressed in terms of the inverse temperature $\beta=\p S/\p H$.

To see whether there is a genuine connection between these zeroth laws, 
restrict from thermodynamics to thermostatics,
in which case the second law of thermodynamics
$$\Delta S\ge\Delta\S\eqno(9.1)$$
becomes an equality.
Here $S$ is the entropy and $\S$ the entropy supply,
as defined in Paper~III.
The classic relation between temperature $\vartheta$, 
entropy flux and thermal flux yields
$$\Delta Q=\vartheta\Delta\S\eqno(9.2)$$ 
in spherical symmetry. 
Alternatively, 
one may regard this as defining absolute temperature $\vartheta$,
as in Gibbsian thermostatics.
Then the quasi-local first law (8.14) implies 
a quasi-local Gibbs equation\footnote*
{Authors less familiar with thermodynamics
often confuse the Gibbs equation, which involves temperature and entropy,
with the first law of thermodynamics, which involves neither.
The two are not equivalent except in thermostatics.}
$$\Delta E=\alpha\vartheta\Delta S-p\Delta V.\eqno(9.3)$$
If $S$ is a function of $V$ and $E$, 
the time derivatives may be replaced with partial derivatives
and {\sl thermal equilibrium} may be defined by constancy of 
$$\beta={\p S\over{\p E}}.\eqno(9.4)$$
Then
$$\beta^{-1}=\alpha\vartheta.\eqno(9.5)$$
Therefore temperature $\vartheta$ is not uniform in thermal equilibrium
in a gravitational field.
This was originally predicted by Tolman [6], 
who argued on different grounds that 
gravitational red-shift leads to a non-uniform temperature.
In the stationary case, $\dot r$ vanishes and $\alpha$ reduces to 
$(1-2E/r)^{1/2}$,
which coincides remarkably with the Tolman red-shift factor.
The above condition is more general 
since it allows thermal equilibrium even in a non-stationary space-time.
Such a possibility is also predicted by 
relativistic kinetic theory of gases [22].

The effective temperature $\alpha\vartheta$ also agrees with that used 
for stationary black holes by Carter [13] and references therein.
Moreover, for the case of Hawking radiation 
for the Schwarzschild black hole, 
the Hawking temperature 
$$\beta^{-1}={\hbar\kappa\over{2\pi}}\eqno(9.6)$$ 
is also an effective temperature,
differing by the same factor from the temperature 
$\vartheta=(1-2E/r)^{-1/2}\beta^{-1}$ 
measured by constant-radius detectors [5].
Here units are such that the Boltzmann constant is unity.
This confirms that active gravitational energy plays the role 
that Gibbsian thermostatics assigns to thermal energy,
strengthening the link between gravity and thermodynamics.

The factor $\kappa\beta=2\pi/\hbar$ 
is not determined by the above argument alone.
However, it does show that if {\sl gravithermal equilibrium}
is defined by a stationary space-time 
with the above condition for thermal equilibrium, 
then both $\kappa$ and $\beta$ are constant, 
which may be interpreted as a weak type of unified zeroth law. 
\bs\ni
{\bf X. Conclusion}
\ms\ni
In summary, the unified first law is just the equation (2.7) expressing 
the gradient of the active gravitational energy 
according to the Einstein equation, divided into energy-supply and work terms.
Projecting this equation along the trapping horizon 
and along the flow of thermodynamic matter yield, respectively, 
first laws of black-hole dynamics (7.1) and relativistic thermodynamics (8.15),
reproduced here:
$$\eqalign{&E'={\kappa A'\over{8\pi}}+wV'\cr
&\dot E=\alpha\dot Q-p\dot V.\cr}$$
In the thermodynamic case, 
this generalises from classical thermodynamics to general relativity.
This shows that active gravitational energy incorporates
thermal energy, as seen explicitly in the Newtonian limit in (8.19),
thereby revealing a physical connection between gravity and thermodynamics.

In the black-hole case, 
this shifts the paradigm from statics to dynamics, 
that is, from stationary to non-stationary black holes.
The standard first law of black-hole statics, 
involving a state-space differential,
has been replaced by a first law of black-hole dynamics, 
involving a horizon derivative.
This is ironically analogous to the paradigm shift 
from thermostatics to thermodynamics that has occured this century,
replacing state-space differentials with material derivatives 
and formulating thermodynamic field equations 
including local first and second laws.
The reader should be warned that this shift to genuine thermodynamics
has been ignored by most authors attempting to draw analogies with black holes,
as well as by most authors of introductory thermodynamics texts.
More enlightened references may be found in Paper~III.

Put another way, the paradigm shift is from Killing horizons and event horizons to trapping horizons.
For instance, Hawking's second law for event horizons is replaced by
the second law of black-hole dynamics of Paper~I, which states that 
the area element of a future outer trapping horizon is non-decreasing,
assuming the null energy condition.
In spherical symmetry this reduces to
$$A'\ge0.$$
Note that the new first and second laws of black-hole dynamics, 
unlike the textbook versions, involve the same derivative. 

Gibbsian definitions of thermal equilibrium and gravithermal equilibrium
have been given, yielding a connection between the zeroth laws,
expressing constancy of surface gravity and an effective temperature.
In contrast, the conventional second law of relativistic thermodynamics
as described in Paper~III, that 
the divergence of the entropy current be non-negative,
is unrelated to the above second law of black-hole dynamics.
Similarly, unlike the case of a quantum field on a stationary background,
there will be no general agreement between the temperature and surface gravity,
since one would expect to describe gravitational collapse of matter 
with arbitrary temperature to black holes with unrelated surface gravity.
Moreover, it is unclear whether the gravitational entropy should agree with 
the area for non-stationary black holes.

The analogies between black-hole dynamics and thermodynamics
have been known to be genuine in only one respect,
namely Hawking's connection between surface gravity and effective temperature
for quantum fields in the stationary case.
This has remained so despite much effort to relate the area of a black hole
to some plausible definition of gravitational entropy 
in general relativity, or indeed, in quite different theories of gravity.
(However, see Ashtekar et al.\ [23] and references therein).
The unified first law has now provided another genuine connection,
neither restricted to the stationary case 
nor dependent on quantum theory.
With hindsight, this may not seem so surprising,
since both textbook first laws effectively express conservation of energy.
However, this was just an analogy in the absence of any obvious link 
between the internal energy of classical thermodynamics 
and the total mass of a black-hole space-time,
and indeed the latter turns out to be not quite right.
The missing link has now been revealed as active gravitational energy.

This provides further evidence for 
the physical importance of gravitational energy.
In particular, 
generalising the results of this paper beyond spherical symmetry would require 
a more general definition of active gravitational energy in general relativity, 
on which there is currently no consensus.
Since energy is also fundamental in quantum theory,
one would expect gravitational energy to be important for quantum gravity.
Moreover, since energy and time share a quantum-mechanical duality,
as expressed by the Heisenberg uncertainty principle,
this is relevant to the oft-discussed problem of time in quantum gravity.
This problem is neatly soluble in spherical symmetry,
where the Kodama vector $k$ encodes a dynamic time, related to $E$.
This suggests that
$$\hat E=i\hbar k\cdot\nabla$$
should be the energy operator of spherically symmetric quantum gravity [18].
Indeed, it may be argued that 
the correspondence principle requires this choice of time.
Along with an area operator such as that provided by the loop theory [23],
a basis for spherically symmetric quantum gravity may thereby be laid.
\bs\ni
{\bf Acknowledgements}\par\ni
Research supported by a European Union Science and Technology Fellowship,
formerly by a Japan Society for the Promotion of Science 
postdoctoral fellowship.
\bs
\begingroup 
\parindent=0pt\everypar={\global\hangindent=20pt\hangafter=1}\par
{\bf References}\ms
\ref{[1] Hayward S A 1994}\PR{D49}{6467}
\ref{[2] Hayward S A 1996}\PR{D53}{1938}
\refb{[3] Hayward S A 1998}{Relativistic thermodynamics}{(gr-qc/9803007)}
\ref{[4] Eckart C 1940}\PR{58}{919}
\refb{[5] Birrell N D \& Davies P C W 1992}{Quantum Fields in Curved Space}
{(Cambridge University Press)}
\refb{[6] Tolman R C 1934}{Relativity, Thermodynamics and Cosmology}
{(Oxford University Press)}
\refb{[7] Hawking S W \& Ellis G F R 1973}
{The Large Scale Structure of Space-Time}{(Cambridge University Press)}
\ref{[8] Misner C W \& Sharp D H 1964}\PR{136}{B571}
\ref{[9] Hayward S A 1994}\PR{D49}{831}
\ref{[10] Hayward S A 1994}\CQG{11}{3037}
\refb{[11] Penrose R \& Rindler W 1988}
{Spinors and Space-Time Volume 2}{(Cambridge University Press)}
\refb{[12] Jackson J D 1962}{Classical Electrodynamics}{(Wiley)}
\refb{[13] Carter B 1973 in}{Black Holes}
{ed.~DeWitt C \& DeWitt B S (Gordon \& Breach)}
\refb{[14] Wald R M 1984}{General Relativity}{(University of Chicago Press)}
\ref{[15] Wald R M 1993}\PR{D48}{R3427}
\ref{[16] Kodama H 1980}\PTP{63}{1217}
\ref{[17] H\'aj\'\i\v cek P 1987}\PR{D36}{1065}
\refb{[18] Hayward S A 1994 in}
{Proc.\ 4th Workshop on General Relativity \& Gravitation}
{ed.\ Nakao K (Kyoto University)}
\ref{[19] Fodor G, Nakamura K, Oshiro Y \& Tomimatsu A 1996}\PR{D54}{3882}
\ref{[20] Visser M 1992}\PR{D46}{2445}
\ref{[21] Heusler M 1995}\CQG{12}{779}
\refb{[22] M\"uller I \& Ruggeri T 1993}{Extended Thermodynamics}
{(Springer-Verlag)}
\ref{[23] Ashtekar A, Baez J, Corichi A \& Krasnov K 1998}\PRL{80}{904}
\endgroup
\bye